\documentclass[11pt]{report}
\usepackage{geometry, graphicx, kotex, imakeidx, titlesec, array} 
\usepackage{mathptmx}   
\usepackage{amsmath, amsthm, amssymb, mathrsfs, multirow, verbatim} 
\usepackage[nobiblatex]{xurl} 
\usepackage{indentfirst} 
\usepackage{booktabs} 
\usepackage{ragged2e}
\usepackage{setspace}
\usepackage{cite}
\usepackage{amsmath,amssymb,amsfonts}
\usepackage{algorithmic}
\usepackage{graphicx}
\usepackage{textcomp}
\usepackage{multirow}
\usepackage{array}
\usepackage{amsmath,amssymb,amsfonts}
\usepackage{graphicx}
\usepackage{textcomp}
\usepackage{multirow}
\usepackage{array}
\usepackage{tikz}
\usepackage{footnote}
\usepackage[ruled,vlined, linesnumbered]{algorithm2e}
\usepackage{rotating}
\usepackage{booktabs}
\usepackage{pdflscape}
\usepackage{longtable}
\usepackage{caption}
\usepackage{float}
\usepackage{rotfloat}

\titleformat{\chapter}{\normalfont\huge\bfseries}{\chaptertitlename\ \thechapter.}{20pt}{\huge} 
\makeindex 

\renewcommand\arraystretch{1.3} 

\begin{document}

\newpage
\thispagestyle{empty}
\begin{center}
\noindent
{\fontsize{16pt}{16pt}\selectfont Doctoral Dissertation \par }  
\vspace{3cm} 
\LARGE Efficient Transformer-Integrated Deep Neural Architectures for Robust EEG Decoding of Complex Visual Imagery 
\par\vspace{4cm} 
{\fontsize{16pt}{16pt}\selectfont Byoung-Hee Kwon \par}
\vspace{0.5cm}
{\fontsize{16pt}{16pt}\selectfont Department of Brain and Cognitive Engineering \par} 
\vspace{1.5cm}
{\fontsize{18pt}{18pt}\selectfont Graduate School \par} 
\vspace{0.5cm}
{\fontsize{18pt}{18pt}\selectfont Korea University \par}
\vspace{1cm}
{\fontsize{14pt}{14pt}\selectfont December 2025} 
\end{center}


\newpage
\thispagestyle{empty}
\begin{center}
\LARGE Efficient Transformer-Integrated Deep Neural Architectures for Robust EEG Decoding of Complex Visual Imagery 
\par\vspace{1.5cm} 
{\fontsize{16pt}{16pt}\selectfont by \par Byoung-Hee Kwon \par} 
\vspace{0.5cm}
\rule{.6\textwidth}{0.4pt} 
\par\vspace{0.2cm}
{\fontsize{16pt}{18pt}\selectfont under the supervision of \par Professor Seong-Whan Lee \par 
\vspace{0.7cm}
A dissertation submitted in partial fulfillment of \par
the requirements for the degree of \par
Doctor of Philosophy \par } 
\vspace{10pt}
{\fontsize{16pt}{16pt}\selectfont Department of Brain and Cognitive Engineering \par }  
\vspace{1.5cm}
{\fontsize{18pt}{18pt}\selectfont Graduate School \par\vspace{0.2cm}
 Korea University \par} 
\vspace{1cm}
{\fontsize{14pt}{14pt}\selectfont December 2025} 
\end{center}


\newpage
\thispagestyle{empty}
\begin{center}
\vspace{1cm}
{\fontsize{16pt}{18pt}\selectfont
The dissertation of Byoung-Hee Kwon has been approved by the dissertation committee in partial fulfillment 
of the requirements for the degree of \par
Doctor of Philosophy \par}
\vspace{1cm}
{\fontsize{14pt}{14pt}\selectfont  December 2025 \par}  
\vspace{2cm}
\rule{.6\textwidth}{0.4pt}\par 
{\fontsize{16pt}{16pt}\selectfont Committee Chair: Seong-Whan Lee \par}
\vspace{1cm}
\rule{.6\textwidth}{0.4pt}\par 
{\fontsize{16pt}{16pt}\selectfont Committee Member: Wonzoo Chung \par}
\vspace{1cm}
\rule{.6\textwidth}{0.4pt}\par 
{\fontsize{16pt}{16pt}\selectfont Committee Member: Tae-Eui Kam \par}
\vspace{1cm}
\rule{.6\textwidth}{0.4pt}\par 
{\fontsize{16pt}{16pt}\selectfont Committee Member: Jichai Jeong \par}
\vspace{1cm}
\rule{.6\textwidth}{0.4pt}\par 
{\fontsize{16pt}{16pt}\selectfont Committee Member: Sanghoon Sull \par}
\vspace{1cm}
\end{center}

\newpage
\pagenumbering{roman} 
\newgeometry{left=20mm, right=20mm, top=30mm, bottom=30mm} 
\begin{center}
\LARGE Efficient Transformer-Integrated Deep Neural Architectures for Robust EEG Decoding of Complex Visual Imagery 
\par\vspace{20pt}

\normalsize \doublespacing
by Byoung-Hee Kwon\par 
Department of Brain and Cognitive Engineering\par
under the supervision of Professor Seong-Whan Lee 
\par\vspace{20pt}
\large \textbf{Abstract}
\end{center}

\normalsize
\justifying 
\doublespacing

This study introduces a pioneering approach in brain-computer interface (BCI) technology, featuring our novel concept of complex visual imagery for non-invasive electroencephalography (EEG)-based communication. Complex visual imagery, as proposed in our work, involves the user engaging in the mental visualization of complex upper limb movements. This innovative approach significantly enhances the BCI system, facilitating the extension of its applications to more sophisticated tasks such as EEG-based robotic arm control. By leveraging this advanced form of visual imagery, our study opens new horizons for intricate and intuitive mind-controlled interfaces. We developed an advanced deep learning architecture that integrates functional connectivity metrics with a convolutional neural network-image transformer. This framework is adept at decoding subtle user intentions, addressing the spatial variability in complex visual tasks, and effectively translating these into precise commands for robotic arm control. Our comprehensive offline and pseudo-online evaluations demonstrate the framework's efficacy in real-time applications, including the nuanced control of robotic arms. The robustness of our approach is further validated through leave-one-subject-out cross-validation, marking a significant step towards versatile, subject-independent BCI applications. This research highlights the transformative impact of advanced visual imagery and deep learning in enhancing the usability and adaptability of BCI systems, particularly in robotic arm manipulation.

\par\vspace{20pt}


\newpage 
\begin{center}
\LARGE 복잡한 시각 심상의 강인한 뇌파(EEG) 디코딩을 위한 효율적인 Transformer 통합 심층 신경망 
\par\vspace{10pt}
\normalsize 권 병 희\par 
뇌 공 학 과\par 
지 도 교 수:  이 성 환
\par\vspace{10pt}
\large \textbf{국문 초록}
\end{center}

\normalsize 

본 연구는 비침습적 뇌전도(EEG) 기반 의사소통을 위한 뇌-컴퓨터 인터페이스(BCI) 기술에서 선구적인 접근법을 소개하며, 복잡한 시각적 심상이라는 새로운 개념을 제시한다. 본 연구에서 제안하는 복잡한 시각적 심상은 사용자가 상지의 복잡한 움직임을 정신적으로 시각화하는 것을 포함한다. 이 혁신적인 접근은 BCI 시스템을 크게 향상시켜, EEG 기반 로봇 팔 제어와 같은 보다 정교한 작업으로 그 응용을 확장할 수 있도록 한다.
이러한 발전된 형태의 시각적 심상을 활용함으로써, 본 연구는 정교하고 직관적인 사고 제어 인터페이스를 위한 새로운 가능성을 제시한다. 우리는 기능적 연결성 지표를 합성곱 신경망과 이미지 트랜스포머와 통합한 고급 딥러닝 아키텍처를 개발하였다. 이 프레임워크는 사용자의 미세한 의도를 디코딩하고, 복잡한 시각 과제에서의 공간적 변동성을 처리하며, 이를 로봇 팔 제어를 위한 정확한 명령으로 효과적으로 변환한다.
포괄적인 오프라인 및 준실시간 평가를 통해 본 프레임워크의 실시간 응용에서의 효율성이 입증되었으며, 로봇 팔의 세밀한 제어에서도 그 성능이 확인되었다. 또한 leave-one-subject-out 교차검증을 통해 본 접근법의 견고성이 추가로 검증되었으며, 이는 사용자 독립적 BCI 응용으로의 중요한 진전을 의미한다.
본 연구는 고급 시각적 심상과 딥러닝이 BCI 시스템의 사용성과 적응성을 향상시키는 데 미치는 혁신적 영향을 강조하며, 특히 로봇 팔 조작에서 그 잠재력을 보여준다.

\par \vspace{10pt}
\textbf{중심어}: 뇌-컴퓨터 인터페이스, 뇌 신호, Transformer, 시각 상상

\newpage
\chapter*{Preface}
\normalsize

This dissertation is submitted for the degree of Doctor of Philosophy in the Department of Brain and Cognitive Engineering at Korea University. The research presented herein was conducted under the supervision of Professor Seong-Whan Lee in the Department of Artificial Intelligence at Korea University. The core content of this dissertation is based on the study published in \textit{Expert Systems with Applications}, volume 294, 2025. Neither this dissertation, nor any substantially similar work, has been or is being submitted for any other degree, diploma, or academic qualification at any other university.

\newpage
\chapter*{Acknowledgment}

This work was partly supported by Institute of Information \& Communications Technology Planning \& Evaluation (IITP) grant funded by the Korea government (MSIT) (No. 2019-0-00079, Artificial Intelligence Graduate School Program(Korea University)) and the National Research Foundation of Korea (NRF) grant funded by the MSIT (No. 2022-2-00975, MetaSkin: Developing Next-generation Neurohaptic Interface Technology that enables Communication and Control in Metaverse by Skin Touch.)

\renewcommand*\contentsname{Contents}
\tableofcontents



\listoffigures


\listoftables




\chapter{Introduction}\label{chap:introduction}
\pagenumbering{arabic} 
Brain-computer interfaces (BCIs) are an advanced technology that enables communication between machines and users, translating human brain signals into machine commands to reflect the user's intentions. Non-invasive BCI is a practical technology as no surgical operation is required \cite{kwak2019error, jeong2022subject, sun2024efficient}. Electroencephalography (EEG) signals have the advantage of having better time resolution than comparable methods such as functional magnetic resonance imaging and near-infrared spectroscopy. Thus, EEG allows rapid communication between users and external devices which facilitates the technological improvement of rehabilitation systems for patients with tetraplegia or supporting the daily life activities of healthy people \cite{suk2012novel, zhou2023shared, ahmad2006human,  ai2023bci}. Using visual imagery, which is a type of intuitive BCI paradigm, users could control their avatars and robotic devices in virtual space and in the real-world environment by reflecting user intentions. Furthermore, it allows the creation of a new type of world in which communication in a 3D virtual space, without any spatial or physical restrictions, becomes possible by using only brain signals.

Previous BCI studies have focused on a variety of BCI paradigms to identify users' intent to move. Among the endogenous BCI paradigms to control BCI-based devices, motor imagery (MI) \cite{sun2019advanced, penaloza2018bmi, jeong2022subject, liu2024cross} is an effective option for controlling practical BCI applications such as robotic arms or wheelchairs. When a user performs MI, event-related desynchronization/synchronization (ERD/ERS), referred to as sensorimotor rhythm, is generated in the motor cortex \cite{jeong2020brain, wang2020common, luo2024selective}. ERD/ERS can be induced from the mu band ([8–12] Hz) and beta band ([13–30] Hz), respectively.
However, despite its prevalence, the MI paradigm in BCI studies poses challenges in intuitiveness for users, especially when they imagine muscle movements. Although a user imagines the same muscle movement when performing the MI paradigm, it is difficult to detect the user’s intentions because the time and sequence of imagery in which the user imagines muscle movements are not constant. In addition, performing the MI task is accepted differently by users, and this phenomenon prevents users from having a uniform imagination \cite{perez2022eegsym, tibrewal2022classification}. This leads to motion-related signals in different brain regions for each person and degrades the practicality of BCI applications.

Visual imagery is an endogenous BCI paradigm that allows users to use more intuitive imagination than MI when performing imagery tasks \cite{kwon2020decoding}, regardless of the complexity of the movement, reducing the difficulty of imagination and minimizing user fatigue. Additionally, it can be used to lead users to perform the visual imagery task uniformly. This effective BCI paradigm presents the possibility for the development of technologies that control avatars or robots by reflecting users' intentions in 3D virtual reality and in the real-world by using only brain signals. To decode user intention, we used neural response patterns that included delta, theta, and alpha frequency bands. Brain activities based on visual imagery induce delta and theta bands in the prefrontal lobe and the alpha band in the occipital lobe, which includes the visual cortex \cite{pearson2019human, koizumi2019eeg}. Previous studies proved that the brain signals generated by visual imagery can be analyzed by specific brain regions and frequencies \cite{sousa2017pure} to control BCI-related devices.

However, most previous visual imagery paradigms for EEG-based BCI have focused on tasks designed in two-dimensional space, often lacking contextual relevance or real-world applicability. Common examples include imagining directional arrows pointing left or right \cite{sousa2017pure}, mentally visualizing geometric shapes such as circles or triangles \cite{llorella2021classification}, or recalling specific objects or colors \cite{xie2020visual, kilmarx2024evaluating}. While these tasks are easy to implement and elicit measurable brain responses, they often involve low task complexity and offer limited relevance to real-world BCI applications. Even in studies that extended visual imagery into three-dimensional space, the tasks were frequently limited to imagining spatial directions or simple trajectories \cite{koizumi2019eeg}, rather than simulating functional, goal-directed actions. As a result, such paradigms may not effectively support intuitive and naturalistic control of external devices. These limitations highlight the need for a more realistic and complex paradigm that better engages the brain’s visual and motor-related regions through structured and contextually grounded imagery tasks. Therefore, this study proposes a complex visual imagery paradigm that not only provides intuitive visual stimuli for users but also delivers complex movement cues required in virtual reality and real-world environments. In addition, it allowes the training of a user by providing repeated 3D space-based visual stimuli. Complex visual imagery allows users to perform more advanced tasks when performing BCI-based device control because the user is given a moving stimulus.

In this study, visual imagery \cite{sousa2017pure, koizumi2019eeg} was used to provide more intuitive stimuli to users. We provided complex visual imagery by which the user imagined complex movements, unlike traditional visual imagery by which the user imagined simple movements. Traditional visual imagery provides simple stimulation for imagining, such as static images or direction indications \cite{kwon2020decoding}. We provided visual stimuli in a three-dimensional space on the monitor to enable users to perform visual imagery effectively. The complex visual imagery paradigm that we designed presented the high-complexity movement stimulation required in virtual reality and real-life to help the user intuitively perform imagery tasks. In addition, it allowed the training of a user by providing repeated 3D space-based visual stimuli. Complex visual imagery allows users to perform more advanced tasks when performing BCI-based device control because the user is given a moving stimulus.

This advanced form of mental imagery, involving the imagination of three-dimensional constructs, engages various brain regions in a sophisticated and coordinated manner. Initially, the occipital lobe, located in the posterior part of the brain, is essential in providing fundamental visual processing capabilities. Although the task is imaginative rather than perceptive, this region still plays a role in offering a base from previously acquired visual memories and experiences, crucial for constructing the initial framework of the imagined 3D image.

The frontal lobe, particularly the dorsolateral prefrontal cortex, then takes a leading role in the cognitive aspects of imagining these 3D constructs. It goes beyond mere visual processing to engage in higher-order cognitive functions. This includes the manipulation of abstract spatial concepts and the organization of complex visual information derived from memory and experience. The frontal lobe's executive functions are vital in this context, enabling the individual to mentally rotate, scale, and manipulate the imagined three-dimensional objects. Additionally, its working memory plays a crucial role in maintaining a dynamic and coherent representation of these imagined constructs over time, allowing for the creation of sustained, detailed, and vivid mental imagery. In summary, while the occipital lobe provides the necessary visual references from memory, the frontal lobe's contribution is pivotal in transforming these references into intricate three-dimensional mental constructs. This process involves complex cognitive abilities like spatial reasoning, memory retrieval, and the imaginative manipulation of visual elements, underscoring the brain's remarkable capacity to not only perceive the external world but also to internally reconstruct and explore complex spatial environments within the mind's eye.

We proposed a deep learning framework based on functional connectivity and a convolutional neural network (CNN)-image transformer network to facilitate robust decoding of complex visual imagery. We decoded brain signals derived from users when imagining four types of motions (e.g., picking up a cell phone, pouring water, opening a door, and eating food) and evaluated the performance of our network, which considers both spatial and temporal information.

The contributions of this study are threefold. (\textit{i}) We proposed a functional connectivity-based deep neural network to decode complex visual imagery from subjects. As a certain spatial pattern is found in visual imagery, our proposed network emphasizes spatial information based on the phase-locking value (PLV) for robust decoding performance. (\textit{ii}) We measured the model performance through not only offline analysis but also pseudo-online analysis to confirm the feasibility of a real-time scenario for a practical BCI. (\textit{iii}) In addition, to contribute to practical BCI application, the possibility of robust classification in the subject-independent condition was investigated by applying a leave-one-subject-out (LOSO) evaluation.

\chapter{Materials and Methods}\label{chap:materialsmethods}
\section{Participants}\label{sec:Participants}
Fifteen subjects (Sub01-Sub15; aged 20–30 years) who had no neurological disease participated in our experiments. The subjects were asked to avoid anything that could affect their physical and mental condition during the experiment, such as drinking alcohol or using psychotropic drugs, during the day before the experiment. In addition, the participants were instructed to sleep for more than 8 h on the day before the experiment. This study was reviewed and approved by the Institutional Review Board at Korea University (KUIRB-2020-0013-01), and written informed consent was obtained from all participants before the experiment.

\begin{figure*}[t!]
\centering
\includegraphics[width=\textwidth]{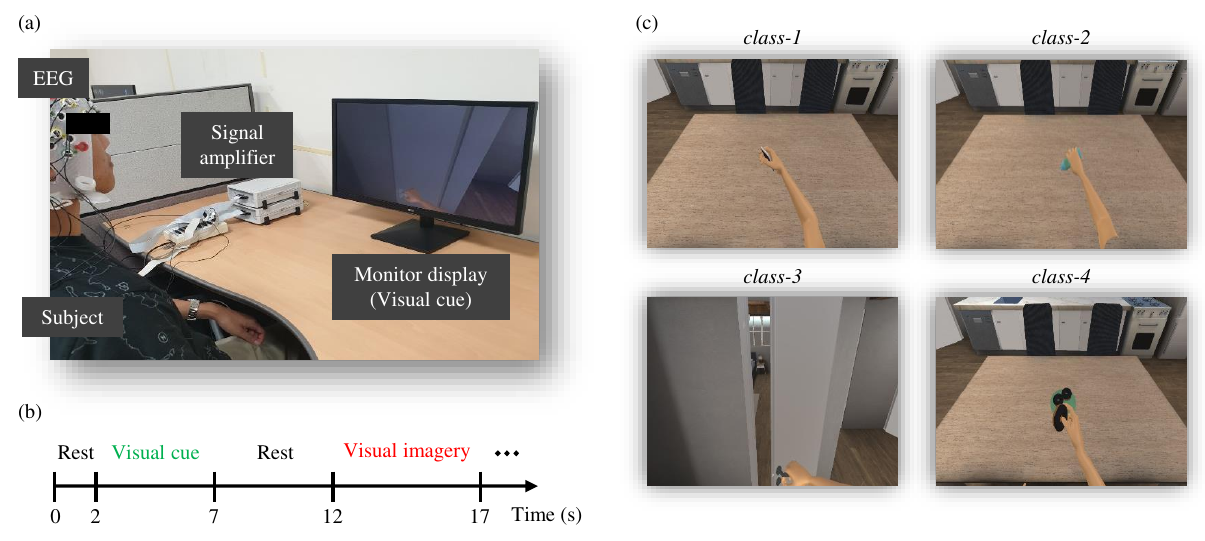}
\caption{Experimental protocols for complex visual imagery from electroencephalography (EEG) signals. (a) Experimental environment for acquiring complex visual imagery data. (b) Experimental paradigm in a single trial and representation of visual cues according to each task. (c) The type of stimuli given to users in the experiment: picking up a cell phone (class-1), pouring water (class-2), opening a door (class-3), and eating food (class-4).}
\end{figure*}

\section{Data Acquisition}\label{sec:Data_Acquisition}
EEG data were recorded using an EEG signal amplifier (BrainAmp, Brain Products GmbH, Gilching, Germany) with MATLAB 2019a (MathWorks, Natick, MA, USA), sampled at 1,000 Hz. Additionally, we applied a 60 Hz notch filter to reduce the effect of external electrical noise (for example, direct-current noise owing to the power supply, scan rate of the monitor display, and frequency of the fluorescent lamp) in the raw signals. EEG was recorded from 64 Ag/AgCl electrodes according to the International 10–20 system (Fp1-2, AF5-6, AF7-8, AFz, F1-8, Fz, FT7-8, FC1-6, T7-8, C1-6, Cz, TP7-8, CP1-6, CPz, P1-8, Pz, PO3-4, PO7-8, POz, O1-2, Oz, and Iz). The ground and reference channels were placed on the Fpz and FCz, respectively. To maintain the impedance between the electrodes and skin below 10 k$\Omega$, we injected conductive gel into the electrodes using a syringe with a blunt needle. A display monitor to provide the experimental paradigm to the subject was placed at a distance of approximately 90 cm to achieve a comfortable state for the subject, and the subject sat in a comfortable position to conduct the experiment, as shown in Fig. 2.1(a).

\section{Experimental Paradigm}\label{sec:protocolparadigm}
As shown in Fig. 2.1(b), in this experimental paradigm, a single trial included four different phases. The first phase was a resting state during which a fixation cross was presented on the screen to provide a comfortable environment for the subjects for 2 s, before providing the visual stimuli. The visual stimuli that subjects should imagine were presented as a visual cue in the second phase. The third phase was another resting state, with the fixation cross presented, to provide sufficient time (5 s) to remove the afterimage of the stimulus presented in the second stage. This stage was essential for obtaining clear and meaningful EEG data from the visual imagery stage. In the fourth stage, after the 5-s resting phase, the subjects performed visual imagery for 5 s based on the visual cues provided during the second stage. During the visual imagery phase, the subjects saw a blank screen, with their eyes open, and imagined drawing a scene on the screen. In this process, the subject was asked to perform visual imagery with the feeling that he/she did not use any muscles, to avoid confusion with the MI. The duration of a single trial consisting of all four above-mentioned stages was 17 s. Each subject performed 50 trials in each class, for a total of 800 trials. Trials for each class were conducted in a random order across subjects.

We designed a training platform with moving visual stimulation, which we called 3D-BCI, as a guide to experimental protocols. These stimuli provide guidance on what the subject should imagine. The subjects performed complex visual imagery based on a visual cue. We used 3D simulation software such as Unity 3D (Unity Technologies, San Francisco, CA, USA) and Blender (Blender 3D Engine: www.blender.org) to design videos that were presented to the subjects. The visual stimuli consisted of four different scenarios: picking up a cell phone (\textit{class-1}), pouring water (\textit{class-2}), opening a door (\textit{class-3}), and eating food (\textit{class-4}). The classes were designed to reflect movements that the user could perform in real life, as well as those required for EEG-based robotic arm control.

\section{Pre-processing}\label{sec:processing}
The EEG data were pre-processed using the BBCI toolbox \cite{blankertz2010berlin} in a MATLAB 2019a environment. Raw EEG data were downsampled from 1,000 Hz to 250 Hz. We applied band-pass filters to extract specific frequency bands significant to visual imagery: delta [0.5-4] Hz, theta [4-8] Hz, and alpha [8-13] Hz, using Hamming-windowed zero-phase finite-impulse response filters with an optimized order ($N$ = 30) \cite{sousa2017pure, lim2000text, koizumi2019eeg}. Opting for an advanced data augmentation technique, we integrated Gaussian noise into our dataset, a method renowned for its efficacy in enhancing data robustness and diversity. Originally, each subject's dataset comprised 800 samples. With the application of our Gaussian noise-based augmentation, we realized a substantial fivefold increase, culminating in 4,000 samples per subject, thereby significantly expanding the training data's dimensional space. With this enriched dataset, the data was systematically partitioned: 60\% served as the foundational training dataset, 20\% was delineated for validation, and the residual 20\% was allocated for testing\cite{lee2020uncertainty, lee1990translation}.

\section{Functional Connectivity-based Deep Neural Network}\label{sec:FCDN}
Functional connectivity is used to assess neural interactions \cite{honey2009predicting}. We utilized functional connectivity as a method to evaluate regional interactions in the brain when a subject performs visual imagery. We selected the PLV method \cite{lachaux1999measuring, roach2008event} as a functional connectivity method to calculate the relation score of each channel.

\begin{figure*}[t!]
\centering
\includegraphics[width=\textwidth]{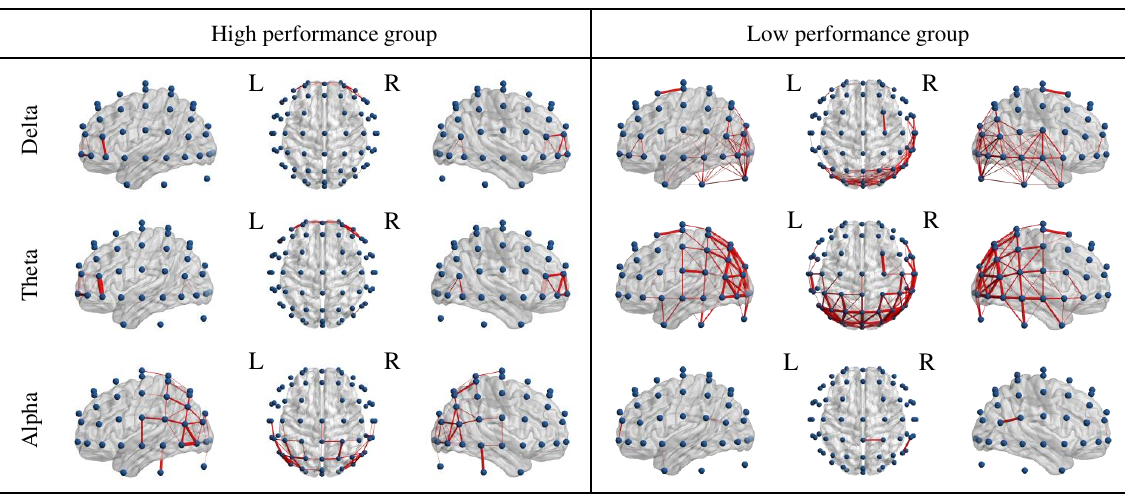}
\caption{The connections between the channels that have functional connectivity scores above 0.9 when the high-performance group (Sub14, Sub09, and Sub04) and low-performance group (Sub01, Sub08, and Sub13) performed a complex visual imagery task. Functional connectivity was assessed in the delta and alpha frequency ranges. In the delta band, connections are mainly represented in the prefrontal area as shown in the high-performance group. On the other hand, the low-performance group's connections are irregular. In the alpha band, the third row, the high-performance group's connections are mainly represented in the occipital area. However, the low-performance group's connections show irregular tendencies.}
\end{figure*}

Brain signals can be expressed through correlations between brain regions. To decode user intention, raw EEG signals, which do not reflect correlations between regions of the brain and the deep neural network, lead to performance degradation. Therefore, we measured the functional connectivity scores using PLVs to consider spatial relationships. As shown in Fig. 2.2, we visualized the functional connectivity when the subjects performed the complex visual imagery in each class, by using chord diagrams to confirm if this method was appropriate for decoding complex visual imagery. In this study, subjects were categorized into high-performance and low-performance groups according to the classification performance criteria detailed in the results section. We identified the functional connectivity in all classes using the delta ([0.5--4] Hz), theta ([4--8] Hz), and alpha ([8--13] Hz) frequency ranges where significant features appear in compl visual imagery. To extract signals from each frequency band, we employed the following generalized formula:

\begin{equation}
    X_{\Delta f}(t) = \int_{f_{\text{min}}}^{f_{\text{max}}} X(f) e^{2\pi i f t} \, df
\end{equation}

In this formula, $X_{\Delta f}(t)$ represents the signal in the time domain for a specific frequency band, $f_{\text{min}}$ and $f_{\text{max}}$ define the lower and upper limits of the frequency band, and $X(f)$ is the frequency spectrum of the original EEG signal. In the high-performance group (Sub14, Sub09, and Sub04), strong correlations were observed primarily near the prefrontal lobe in the delta and theta regions, and in the alpha frequency range, predominantly near the occipital lobe. 

On the other hand, the low-performance group (Sub01, Sub08, and Sub13) demonstrated weak connections in the prefrontal lobe and occipital lobe. These results demonstrated that applying PLVs to decode complex visual imagery is reasonable.

\begin{figure*}[t]
\centering
\includegraphics[width=\textwidth]{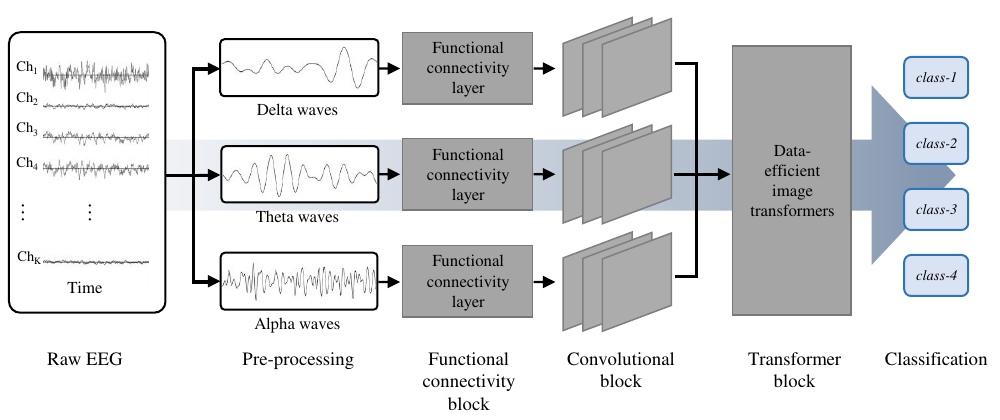}
\caption{The overview of the proposed architecture begins with the segmentation of the raw electroencephalogram (EEG) into delta waves, theta waves, and alpha waves. In the functional connectivity layer, each of these EEG wave types is then multiplied channel-wise using a functional connectivity score measured by the phase-locking value. The modified EEG data that emphasize spatial information is used as the input data for the deep learning architecture. We extracted temporal information using a network consisting of a convolutional block, followed by spatial information via a transformer block.
}
\end{figure*}

\begin{figure}[]
\centering
\includegraphics[width=0.75\textwidth]{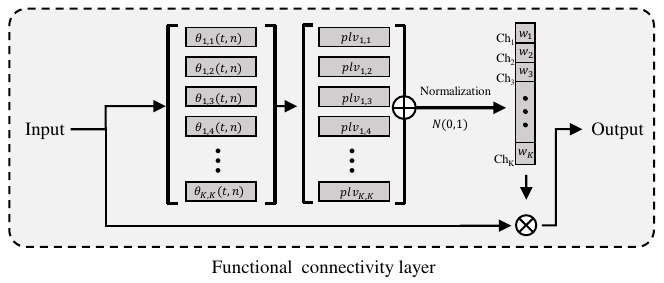}
\caption{In the functional connectivity layer, the raw electroencephalogram (EEG) data undergo transformation through the phase locking value (PLV). Each channel's connectivity score, derived via the PLV, is then normalized. The subsequent integration with the original EEG accentuates spatial details, producing an EEG output enriched with heightened spatial significance.
}
\end{figure}

\begin{table}[t!]
\caption{Description of the Proposed Architecture}
\begin{center}
\resizebox{0.8\textwidth}{!}{%
\begin{tabular}{cccc}
\hline
\textbf{Layer}     & \textbf{Type}   & \textbf{Parameter}                                                                                          & \textbf{Output size}                      \\ \hline
1                  & Input           & -                                                                                                           & 1$\times$64$\times$1000                   \\ \hline
\multirow{2}{*}{2} & Convolution     & \begin{tabular}[c]{@{}c@{}}Filter size:1$\times$20\\ Stride size: 1$\times$1\\ Feature map: 40\end{tabular} & \multirow{2}{*}{40$\times$64$\times$981}  \\ \cline{2-3}
                   & BatchNorm       & -                                                                                                           &                                           \\ \hline
\multirow{3}{*}{3} & Convolution     & \begin{tabular}[c]{@{}c@{}}Filter size:1$\times$20\\ Stride size: 1$\times$1\\ Feature map: 40\end{tabular} & \multirow{3}{*}{80$\times$64$\times$962}  \\ \cline{2-3}
                   & BatchNorm       & -                                                                                                           &                                           \\ \cline{2-3}
                   & Activation(ELU) & -                                                                                                           &                                           \\ \hline
\multirow{3}{*}{4} & Convolution     & \begin{tabular}[c]{@{}c@{}}Filter size:1$\times$40\\ Stride size: 1$\times$1\\ Feature map: 80\end{tabular} & \multirow{3}{*}{160$\times$64$\times$962} \\ \cline{2-3}
                   & BatchNorm       & -                                                                                                           &                                           \\ \cline{2-3}
                   & Activation(ELU) & -                                                                                                           &                                           \\ \hline
\multirow{2}{*}{5} & Average pooling & \begin{tabular}[c]{@{}c@{}}Filter size: 1$\times$32\\ Stride size: 1$\times$32\\ Dropout(0.5)\end{tabular}            & \multirow{2}{*}{160$\times$64$\times$30}  \\ \cline{2-3}
                   & Activation(ELU) & -                                                                                                           &                                           \\ \hline
6                  & Average pooling & \begin{tabular}[c]{@{}c@{}}Filter size: 1$\times$30\\ Stride size: 1$\times$30\\ Dropout(0.5)\end{tabular}            & 160$\times$64$\times$1                    \\ \hline
7                  & Interpolation   & -                                                                                                           & 224$\times$224                                 \\ \hline
8                  & Concatenation & -                                                                                                           & 224$\times$224$\times$3                               \\ \hline
9                  & DeiT            & -                                                                                                           & 1$\times$4                                     \\ \hline
\end{tabular}
}
\end{center}
\end{table}

In this study, we proposed a functional connectivity-guided deep neural network (FCDN) to decode acquired complex visual imagery data. The overall processing pipeline of the proposed FCDN begins with frequency-wise decomposition of the raw EEG signals, followed by spatial enhancement using functional connectivity scores. The resulting signals are then passed through a convolutional block to extract temporal features, which are subsequently transformed into spatial feature maps and processed by a data-efficient transformer for final classification. An overview of the proposed FCDN is presented in Fig. 2.3, and the architecture design of the deep learning part is presented in Table 2.1. As a first step of the proposed network, raw EEG signals were decomposed into distinct frequency bands, specifically delta, theta, and alpha. These segregated bands were then utilized as inputs to their respective functional connectivity layers, serving as an input to the subsequent convolutional processing. A detailed representation of the functional connectivity layer is illustrated in Fig. 2.4, where the specific format of this layer is mathematically denoted as

\begin{equation}
PLV=\left \{ plv_{k_1,k_2} \mid 1 \le k_{1} \le K, 1 \le k_{2} \le K \right \},
\end{equation}

\noindent 
where $k_{1}$ and $k_{2}$ are the channels in which we compare the phase, and $K$ is the number of electrodes used. The phase signal $\phi_{k_{1}}(t,n)$ is extracted for all time bins $t$ and trials $n$ $[1, ... , N]$, and the phase difference between $k_{1}$ and $k_{2}$ is derived from $ \theta(t,n)=\phi_{k_{1}}(t,n)-\phi_{k_{2}}(t,n)$.

\begin{equation}
plv_{k_{1},k_{2}}= \frac{1}{NT} \sum_{t=1}^T \sum_{n=1}^N e^{j\theta (t,n)}
\end{equation}

These formulas can be used to understand the correlation between channels and to emphasize the spatial information in raw EEG signals. $plv_{k_{1},k_{2}}$ obtained from the above formula is an upper triangular matrix, and it was transposed to create a lower triangular matrix and added to the two matrices to create a new matrix $S_{k_{1} and, k_{2}}$ in the form of a symmetric matrix.

\begin{equation}
    S_{k_{1},k_{2}}=plv_{k_{1},k_{2}} + (plv_{k_{1},k_{2}})^T
\end{equation}

To emphasize the spatial information in the input signal, we transformed the previously obtained $S_{k_{1},k_{2}}$ into a matrix of 1 $\times$ $K$, using the following formula: 

\begin{equation}
    \widetilde{PLV} = \sum_{k_{1}=1}^K S_{k_{1},k_{2}}
\end{equation}

We limited the distribution of $\widetilde{PLV}$ between 0 and 1 using the simplest method, min-max normalization, to emphasize channels using $\widetilde{PLV}$ implemented in one dimension.

\begin{equation}
    w_{K} = {\widetilde{PLV}_{K} - \widetilde{PLV}_{min} \over \widetilde{PLV}_{max} - \widetilde{PLV}_{min}}. 
\end{equation}

Consequently, for the 64 channels, the values of the channels that were highly correlated with the visual imagery approached 1, while that of channels that were not highly correlated with the complex visual imagery had values that approached zero. $w_{K}$ is a weight that emphasizes spatial information using channels related to visual imagery in the proposed network. The channel axis of the pre-processed EEG signal is multiplied by $w_{K}$, and the data are reconstructed according to the degree of visual imagery.

Conventional EEG-related deep learning frameworks use CNNs to train temporal features \cite{zhang2018spatial, amin2019deep, maeng2012nighttime, zhang2019making} and use image transformers to train spatial features \cite{roh2007accurate, mulkey2023supervised, hwang2006full}. Based on these studies, we adopted the CNN framework in our proposed network to extract the information of spatially focused data using PLVs. We trained the network for 200 epochs, the batch size of the training process was set to 16, and the learning rate was set at 0.0001. When features are extracted from reconstructed EEG data using PLVs, the effect of emphasizing the channel that is most affected when performing visual imagery is greater than that of extracting features from raw EEG data.

The convolution block was designed to handle all electrode channels and to consider the temporal information of the reconstructed data. During this stage, each layer extracted the information that contained temporal information. Batch normalization was applied to standardize the variance of the learned data. Because the first convolutional layer did not improve performance even with nonlinear activation\cite{yang2007reconstruction}, we designed the network to maintain linearity without using the activation function in the first layer. In the second and third convolutional layers, we applied exponential linear units (ELUs) \cite{clevert2015fast, min2022attentional} as an activation function. After the three convolutional layers, two average pooling layers were applied to down-sample the feature maps, primarily focusing on the extraction of temporal information. This restructuring to a 2D format was critical for subsequent integration with the data-efficient image transformers (DeiT) \cite{touvron2021training}, facilitating optimal capture of spatial information. In addition, we set the dropout probability to 0.5, after each average pooling layer, to help prevent the overfitting problem that occurred during training on small sample sizes. Then, we used bicubic interpolation \cite{dong2015image, lee2001automatic} to resize the output of the convolution block.

\begin{align}
    F_{bicubic}(i,j) = \sum_{m=-1}^2 \sum_{n=-1}^2{a_{mn} \cdot (i-x_{0})^{m} \cdot (j-y_{0})^{n}}
\end{align}
where $F_{bicubic}$ is the feature map obtained after applying bicubic interpolation, where $i$ and $j$ denote the indices of rows and columns in the new feature map, respectively. The variables $x_{0}$ and $y_{0}$ correspond to the row and column indices of the original feature map, and a signifies the coefficients used in bicubic interpolation, computed using the neighboring pixel values of the corresponding pixel in the original image. These coefficients must be pre-computed, as they are employed to fit a cubic polynomial. Additionally, $m$ and $n$ are indices ranging from -1 to 2, representing a 4$\times$4 pixel block. This facilitates utilization of the 16 pixels surrounding each pixel, to compute the new pixel value in the interpolated image.

After the bicubic interpolation, features derived from the distinct frequency bands—delta, theta, and alpha—are resized to a dimension of 224$\times$224. Prior to further processing, these values are systematically normalized to fit within a 0 to 255 scale. This normalization step ensures the resultant feature map aligns with typical RGB channel conventions, thereby preparing the data for optimal compatibility with the subsequent deep learning architecture. Following this normalization, features from the three frequency domains are concatenated, resulting in a combined feature map of 224$\times$224$\times$3. This procedure yields a comprehensive representation, integrating spectral EEG characteristics across the examined frequency bands. This unified 224$\times$224$\times$3 feature map is then input into the DeiT. Within the DeiT architecture, the combined EEG features are subjected to a series of convolutional and transformer-based operations, allowing for intricate spatial pattern recognition and further refinement.

In the realm of EEG signal processing, the scarcity of available data often poses significant challenges to conventional deep learning models. This limitation hampers the ability to effectively train large-scale models, thereby restricting the extraction of intricate and subtle features inherent in EEG signals. To mitigate this problem, our approach leverages the DeiT, a model specifically tailored to perform well with limited data.

The cornerstone of DeiT that makes it apt for handling limited data is the employment of a teacher-student distillation process. The key equation governing this distillation is as follows:

\begin{align}
   L_{distill} = \alpha \cdot L_{cls} + \beta \cdot \sum_{i=1}^N{[sim(T_{i},S_{i})\cdot q_{i}]}
\end{align}

where $L_{distill}$ is the total distillation loss, $L_{cls}$ is the classification loss, $\alpha$ and $\beta$ are hyperparameters controlling the balance between the two terms, $T_{i}$ and $S_{i}$ are the teacher's and student's hidden representations at layer $i$, $sim(\cdot,\cdot)$ is the similarity function, and $q_i$ is the corresponding student's attention map.

By leveraging the distinctive information present in the EEG data's delta, theta, and alpha waves, we construct three individual feature maps of dimension 224$\times$224. Each of these maps has been created by undergoing processing through a functional connectivity block and a convolution block, with intricate features representative of their respective waveforms being extracted in the process. By stacking these processed maps along a third dimension, we obtain a composite representation of shape 224$\times$224$\times$3. This method allows for an integrated perspective of the EEG signals, encapsulating the significant information and interactions of the delta, theta, and alpha frequencies.

Such a concatenated representation is seamlessly integrated with the DeiT model. This model architecture, which orchestrates a similarity loss between the hidden representations of a substantial pre-trained teacher model and a more compact student counterpart, guides the student to emulate the teacher's behavior. This alignment capitalizes on the comprehensive knowledge contained within the teacher model, eliminating the necessity for extensive data. Furthermore, this mechanism enables DeiT to discern patterns across different regions of the 224$\times$224$\times$3 feature maps, adeptly harnessing spatial relationships and thereby achieving nuanced image understanding. The inclusion of a self-attention formulation further ensures the capture of complex dependencies, which empowers our model for this specific task. As a concluding step, the extracted features are introduced to a softmax classifier, with the categorical cross-entropy loss function and the Adam optimizer employed for training purposes.


\chapter{Experimental Results}\label{chap:results}
\section{Data Analysis}\label{sec:Dataanalysis}

We used the BBCI toolbox and EEGLab toolbox \cite{delorme2004eeglab} (version 14.1.2b) to verify the quality of the collected EEG data. The BBCI toolbox was used to measure the magnitude of the power of the complex visual imagery data for each frequency range in each area of the brain. We used complex visual imagery data for each stimulus and adopted a period of 0$\sim$5 s for the entire interval of the visual imagery phase. Among the 64 channels, Fz, representing the prefrontal lobe, and Oz, representing the occipital lobe, were selected to measure the power spectral changes \cite{wang2012translation, carrier2001effects} from 0.1 Hz to 60 Hz in each selected channel. The frequency-specific power measured at each channel determines the area of the brain, which is significant when performing complex visual imagery. While the user performed imagery, brain signal power changed significantly at channels representing the prefrontal and occipital lobes. A delta power peak was observed at the Fz channel (Fig. 3.1(a)), while delta power and alpha power peaks were found at the Oz channel (Fig. 3.1(b)).

\begin{figure}[t!]
\centering
\includegraphics[width=0.7\textwidth]{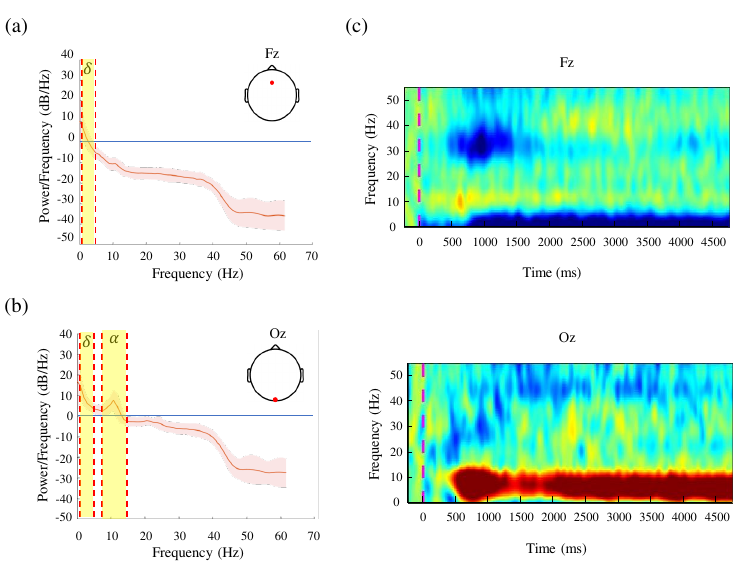}
\caption{The power spectral changes in each selected channel. The yellow box in the graph indicates the frequency range containing a significant peak. The peaks were observed in channels (a) Fz and (b) Oz related to visual imagery. The time-frequency analysis (c) presents the distinct characteristics of complex visual imagery. Prominent features are evident in the alpha band at the Fz and Oz locations, which are known to be associated with visual imagery.}
\end{figure}

Subsequently, we measured the variation in the spectral power of visual imagery based on the event-related spectral perturbation (ERSP) method \cite{makeig1993auditory, delorme2004eeglab} on two channels that are considered to be related to visual imagery. ERSP analysis was performed between 0.5-50 Hz, using 400 time points. The baseline was set from -500 to 0 ms before the visual imagery phase in order to analyze the phenomena when subjects imagined an action. From the above results, it can be inferred that complex visual imagery exhibits strong brain-signal features in the prefrontal and occipital lobes. Accordingly, we measured ERSP on Fz and Oz, two channels related to the visual imagery mentioned above. As shown in Fig. 5(c), Fz, which represents the prefrontal lobe, shows a strong activity power spectrum in the delta wave, and Oz, which represents the occipital lobe, shows a strong activity power spectrum in the alpha wave.

\section{Comparison of Classification Performance using FCDN and Baseline Methods}\label{sec:neurophysiological}

\begin{table}[]
\caption{Comparison of Visual Imagery Classification Performances with Conventional Methods}
\begin{center}
\small{
\renewcommand{\arraystretch}{1.1}
\resizebox{\columnwidth}{!}{%
\begin{tabular}{ccccccc}
\hline
\multicolumn{1}{l}{} & \multicolumn{6}{c}{\textbf{Method}}                                                                                                                                                                                                                                                                                                                                          \\ \hline
\textbf{Subject}     & FBCSP                                                      & EEGNet                                                     & ConvNet                                                    & FBCNet                                                      & TSformer                                              & FCDN                                                       \\ \hline
Sub01                & \begin{tabular}[c]{@{}c@{}}0.4395\\ (±0.0127)\end{tabular} & \begin{tabular}[c]{@{}c@{}}0.5400\\ (+0.0252)\end{tabular} & \begin{tabular}[c]{@{}c@{}}0.5650\\ (+0.0129)\end{tabular} & \begin{tabular}[c]{@{}c@{}}0.5664\\ (±0.0154)\end{tabular}  & \begin{tabular}[c]{@{}c@{}}0.6200\\ (±0.0260)\end{tabular} & \begin{tabular}[c]{@{}c@{}}\textbf{0.6273}\\ \textbf{(±0.0123)}\end{tabular} \\
Sub02                & \begin{tabular}[c]{@{}c@{}}0.4115\\ (±0.0209)\end{tabular} & \begin{tabular}[c]{@{}c@{}}0.5625\\ (+0.0362)\end{tabular} & \begin{tabular}[c]{@{}c@{}}0.6113\\ (+0.0529)\end{tabular} & \begin{tabular}[c]{@{}c@{}}0.6143\\ (±0.0246)\end{tabular} & \begin{tabular}[c]{@{}c@{}}0.6639\\ (±0.0218)\end{tabular} & \begin{tabular}[c]{@{}c@{}}\textbf{0.7135}\\ \textbf{(±0.0195)}\end{tabular} \\
Sub03                & \begin{tabular}[c]{@{}c@{}}0.4013\\ (±0.0173)\end{tabular} & \begin{tabular}[c]{@{}c@{}}0.5988\\ (+0.0195)\end{tabular} & \begin{tabular}[c]{@{}c@{}}0.6248\\ (+0.0276)\end{tabular} & \begin{tabular}[c]{@{}c@{}}0.6386\\ (±0.0152 )\end{tabular} & \begin{tabular}[c]{@{}c@{}}\textbf{0.6894}\\ \textbf{(±0.0179)}\end{tabular} & \begin{tabular}[c]{@{}c@{}}0.6738\\ (±0.0207)\end{tabular} \\
Sub04                & \begin{tabular}[c]{@{}c@{}}0.5303\\ (±0.0105)\end{tabular} & \begin{tabular}[c]{@{}c@{}}0.6900\\ (±0.0412)\end{tabular} & \begin{tabular}[c]{@{}c@{}}0.6888\\ (±0.0266)\end{tabular} & \begin{tabular}[c]{@{}c@{}}0.7121\\ (±0.0148 )\end{tabular} & \begin{tabular}[c]{@{}c@{}}0.7665\\ (±0.0136)\end{tabular} & \begin{tabular}[c]{@{}c@{}}\textbf{0.7701}\\ \textbf{(±0.0214)}\end{tabular} \\
Sub05                & \begin{tabular}[c]{@{}c@{}}0.5982\\ (±0.0216)\end{tabular} & \begin{tabular}[c]{@{}c@{}}0.5800\\ (±0.0139)\end{tabular} & \begin{tabular}[c]{@{}c@{}}0.6038\\ (±0.0404)\end{tabular} & \begin{tabular}[c]{@{}c@{}}0.6428\\ (±0.0250)\end{tabular}  & \begin{tabular}[c]{@{}c@{}}\textbf{0.6938}\\ \textbf{(±0.0163)}\end{tabular} & \begin{tabular}[c]{@{}c@{}}0.6871\\ (±0.0186)\end{tabular} \\
Sub06                & \begin{tabular}[c]{@{}c@{}}0.5939\\ (±0.0298)\end{tabular} & \begin{tabular}[c]{@{}c@{}}0.5850\\ (±0.0396)\end{tabular} & \begin{tabular}[c]{@{}c@{}}0.6100\\ (±0.0252)\end{tabular} & \begin{tabular}[c]{@{}c@{}}0.6537\\ (±0.0229)\end{tabular}  & \begin{tabular}[c]{@{}c@{}}0.7052\\ (±0.0212)\end{tabular} & \begin{tabular}[c]{@{}c@{}}\textbf{0.7203}\\ \textbf{(±0.0188)}\end{tabular} \\
Sub07                & \begin{tabular}[c]{@{}c@{}}0.5205\\ (±0.0237)\end{tabular} & \begin{tabular}[c]{@{}c@{}}0.5888\\ (±0.0266)\end{tabular} & \begin{tabular}[c]{@{}c@{}}0.6175\\ (±0.0195)\end{tabular} & \begin{tabular}[c]{@{}c@{}}0.6142\\ (±0.0231)\end{tabular}  & \begin{tabular}[c]{@{}c@{}}0.6638\\ (±0.0240)\end{tabular} & \begin{tabular}[c]{@{}c@{}}\textbf{0.6813}\\ \textbf{(±0.0239)}\end{tabular} \\
Sub08                & \begin{tabular}[c]{@{}c@{}}0.4005\\ (±0.0239)\end{tabular} & \begin{tabular}[c]{@{}c@{}}0.5450\\ (±0.0183)\end{tabular} & \begin{tabular}[c]{@{}c@{}}0.5975\\ (±0.0459)\end{tabular} & \begin{tabular}[c]{@{}c@{}}0.6297\\ (±0.0183)\end{tabular}  & \begin{tabular}[c]{@{}c@{}}0.6801\\ (±0.0105)\end{tabular} & \begin{tabular}[c]{@{}c@{}}\textbf{0.7195}\\ \textbf{(±0.0267)}\end{tabular} \\
Sub09                & \begin{tabular}[c]{@{}c@{}}0.3864\\ (±0.0291)\end{tabular} & \begin{tabular}[c]{@{}c@{}}0.7750\\ (±0.0230)\end{tabular} & \begin{tabular}[c]{@{}c@{}}0.7650\\ (±0.0364)\end{tabular} & \begin{tabular}[c]{@{}c@{}}0.6483\\ (±0.0257)\end{tabular}  & \begin{tabular}[c]{@{}c@{}}0.6996\\ (±0.0183)\end{tabular} & \begin{tabular}[c]{@{}c@{}}\textbf{0.8207}\\ \textbf{(±0.0196)}\end{tabular} \\
Sub10                & \begin{tabular}[c]{@{}c@{}}0.4782\\ (±0.0121)\end{tabular} & \begin{tabular}[c]{@{}c@{}}0.5813\\ (±0.0468)\end{tabular} & \begin{tabular}[c]{@{}c@{}}0.6288\\ (±0.0200)\end{tabular} & \begin{tabular}[c]{@{}c@{}}0.6809\\ (±0.0273)\end{tabular}  & \begin{tabular}[c]{@{}c@{}}\textbf{0.7338}\\ \textbf{(±0.0198)}\end{tabular} & \begin{tabular}[c]{@{}c@{}}0.7118\\ (±0.0265)\end{tabular} \\
Sub11                & \begin{tabular}[c]{@{}c@{}}0.5293\\ (±0.0293)\end{tabular} & \begin{tabular}[c]{@{}c@{}}0.5263\\ (±0.0108)\end{tabular} & \begin{tabular}[c]{@{}c@{}}0.5775\\ (±0.0303)\end{tabular} & \begin{tabular}[c]{@{}c@{}}0.6685\\ (±0.0229)\end{tabular}  & \begin{tabular}[c]{@{}c@{}}\textbf{0.7208}\\ \textbf{(±0.0242)}\end{tabular} & \begin{tabular}[c]{@{}c@{}}0.6601\\ (±0.0105)\end{tabular} \\
Sub12                & \begin{tabular}[c]{@{}c@{}}0.4229\\ (±0.0116)\end{tabular} & \begin{tabular}[c]{@{}c@{}}0.6488\\ (±0.0174)\end{tabular} & \begin{tabular}[c]{@{}c@{}}0.6275\\ (±0.0261)\end{tabular} & \begin{tabular}[c]{@{}c@{}}0.6316\\ (±0.0188)\end{tabular}  & \begin{tabular}[c]{@{}c@{}}0.6820\\ (±0.0239)\end{tabular} & \begin{tabular}[c]{@{}c@{}}\textbf{0.6951}\\ \textbf{(±0.0177)}\end{tabular} \\
Sub13                & \begin{tabular}[c]{@{}c@{}}0.5231\\ (±0.0231)\end{tabular} & \begin{tabular}[c]{@{}c@{}}0.5850\\ (±0.0246)\end{tabular} & \begin{tabular}[c]{@{}c@{}}0.5813\\ (±0.0220)\end{tabular} & \begin{tabular}[c]{@{}c@{}}0.6039\\ (±0.0225)\end{tabular}  & \begin{tabular}[c]{@{}c@{}}0.6531\\ (±0.0113)\end{tabular} & \begin{tabular}[c]{@{}c@{}}\textbf{0.7203}\\ \textbf{(±0.0190)}\end{tabular} \\
Sub14                & \begin{tabular}[c]{@{}c@{}}0.4360\\ (±0.0235)\end{tabular} & \begin{tabular}[c]{@{}c@{}}0.7950\\ (±0.0325)\end{tabular} & \begin{tabular}[c]{@{}c@{}}0.8438\\ (±0.0205)\end{tabular} & \begin{tabular}[c]{@{}c@{}}0.8046\\ (±0.0214)\end{tabular}  & \begin{tabular}[c]{@{}c@{}}0.8636\\ (±0.0149)\end{tabular} & \begin{tabular}[c]{@{}c@{}}\textbf{0.8720}\\ \textbf{(±0.0139)}\end{tabular} \\
Sub15                & \begin{tabular}[c]{@{}c@{}}0.5892\\ (±0.0129)\end{tabular} & \begin{tabular}[c]{@{}c@{}}0.7125\\ (±0.0240)\end{tabular} & \begin{tabular}[c]{@{}c@{}}0.7288\\ (±0.0233)\end{tabular} & \begin{tabular}[c]{@{}c@{}}0.7179\\ (±0.0104)\end{tabular}  & \begin{tabular}[c]{@{}c@{}}0.7726\\ (±0.0188)\end{tabular} & \begin{tabular}[c]{@{}c@{}}\textbf{0.7785}\\ \textbf{(±0.0264)}\end{tabular} \\
Avg.                 & \begin{tabular}[c]{@{}c@{}}0.4765\\ (±0.0926)\end{tabular} & \begin{tabular}[c]{@{}c@{}}0.6209\\ (±0.0845)\end{tabular} & \begin{tabular}[c]{@{}c@{}}0.6447\\ (±0.0783)\end{tabular} & \begin{tabular}[c]{@{}c@{}}0.6552\\ (±0.1013)\end{tabular}  & \begin{tabular}[c]{@{}c@{}}0.7072\\ (±0.0645)\end{tabular} & \begin{tabular}[c]{@{}c@{}}\textbf{0.7234}\\ \textbf{(±0.0614)}\end{tabular} \\ \hline
\textit{p}           & \textless{}0.05                                            & \textless{}0.05                                            & \textless{}0.05                                            & \textless{}0.05                                             & \textless{}0.05                                            & \textless{}0.05                                            \\ \hline
\end{tabular}
}}
\end{center}
\end{table}

Table 3.1 compares the classification performance of the FCDN and conventional decoding methods. We performed five-fold cross-validation under a subject-dependent scheme to evaluate the classification performance of the proposed FCDN and baseline methods. We used the filter bank common spatial pattern (FBCSP) \cite{ang2008filter}, EEGNet \cite{lawhern2018eegnet}, ConvNet \cite{schirrmeister2017deep}, FBCNet \cite{mane2021fbcnet}, and TSformer \cite{ahn2022multiscale} as baseline methods to decode complex visual imagery data. The FBCSP methodology proficiently extracts spatial patterns across a spectrum of frequency bands, contributing significantly to the analysis of EEG signals. While EEGNet and ConvNet, as CNN-based models, emphasize local features due to their convolutional layers. While this approach of using convolutional layers in EEGNet and ConvNet has its merits, it might restrict their capability to encompass the wider spectrum of EEG signal characteristics. To address this, FBCNet integrates CNNs with handcrafted features, aiming to balance the representation of EEG signal attributes. Additionally, the TSformer, employing transformer techniques, primarily focuses on temporal dynamics, thereby achieving high performance. This method illustrates the potential of advanced signal processing techniques in enhancing EEG signal analysis. We computed a non-parametric paired permutation test \cite{maris2007nonparametric} to confirm the statistically significant differences between the classification results of the proposed method and the baseline methods. In this study, the average classification performance using the proposed network was 0.7234. Comparative analysis with other models showed that EEGNet achieved a performance of 0.6209, ConvNet reached 0.6447, FBCNet scored 0.6552, and TSformer attained 0.7072.

\section{Ablation Study}
In this study, we proposed FCDN, a functional connectivity-guided deep neural network designed for robust classification of EEG signal variability. To evaluate the contribution of the functional connectivity block, we visualized the output features at each major stage of the network using the t-distributed stochastic neighbor embedding (t-SNE) method \cite{jebelli2018continuously, arsalan2019classification}.

\begin{figure}[t!]
\centering
\includegraphics[width=1\textwidth]{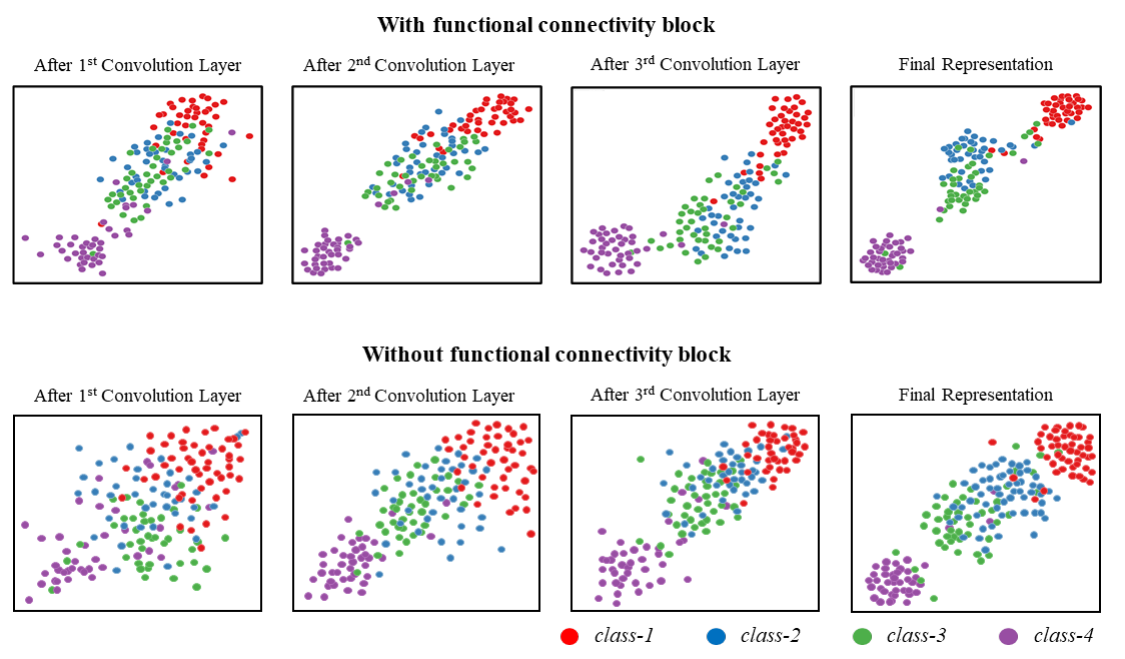}
\caption{Representation of visualized feature distributions in each network layer using t-distributed stochastic neighbor embedding, with and without the functional connectivity block. Each column corresponds to the output after the $1^{st}$, $2^{nd}$, and $3^{rd}$ convolution layers, and the final representation after the DeiT module. The top row shows the full FCDN model with the functional connectivity block, while the bottom row presents the ablation model without the functional connectivity block. As the network deepens, class-wise feature clustering becomes progressively clearer. Notably, the model with the functional connectivity block produces more distinct and compact class clusters compared to the model without the functional connectivity block, demonstrating the critical role of functional connectivity in enhancing spatial discriminability for complex visual imagery decoding.}
\end{figure}

Fig. 3.2 presents the t-SNE distributions of test-set features extracted after the $1^{st}$, $2^{nd}$, and $3^{rd}$ convolutional layers as well as the final representation from the DeiT module. The top row illustrates the feature representations obtained from the full FCDN model that includes the functional connectivity block, whereas the bottom row shows the results from the ablation model in which the functional connectivity block is removed. This structure allows for a direct visual comparison of the impact of the functional connectivity block across network depth. As the network becomes deeper, the feature clustering becomes more distinct. Notably, the model with the functional connectivity block produces tighter and more separable class clusters compared to the model without it. These observations suggest that the functional connectivity block enhances the spatial discriminability of neural representations during complex visual imagery decoding.

\begin{table}[t!]
\caption{Comparison of Classification Accuracies With and Without the Functional Connectivity Block}
\begin{center}
\renewcommand{\arraystretch}{1.3}
\begin{tabular}{c c c c}
\hline
\textbf{Subject} 
& \makebox[3.2cm][c]{\textbf{With}} 
& \makebox[3.2cm][c]{\textbf{Without}} 
& \makebox[3.2cm][c]{\textbf{$\Delta$ Accuracy}} \\ \hline
Sub01 & 0.6273 & 0.6005 & 0.0268 \\
Sub02 & 0.7135 & 0.6571 & 0.0564 \\
Sub03 & 0.6738 & 0.6427 & 0.0311 \\
Sub04 & 0.7701 & 0.7652 & 0.0049 \\
Sub05 & 0.6871 & 0.6575 & 0.0296 \\
Sub06 & 0.7203 & 0.6996 & 0.0207 \\
Sub07 & 0.6813 & 0.6554 & 0.0259 \\
Sub08 & 0.7195 & 0.7048 & 0.0147 \\
Sub09 & 0.8207 & 0.7911 & 0.0296 \\
Sub10 & 0.7118 & 0.6859 & 0.0259 \\
Sub11 & 0.6601 & 0.6381 & 0.0220 \\
Sub12 & 0.6951 & 0.7037 & -0.0086 \\
Sub13 & 0.7203 & 0.6845 & 0.0358 \\
Sub14 & 0.8720 & 0.8467 & 0.0253 \\
Sub15 & 0.7758 & 0.7623 & 0.0135 \\ \hline
Avg.  & 0.7234 & 0.6997 & 0.0237 \\ \hline
\end{tabular}
\end{center}
\end{table}


\begin{figure}[t!]
\centering
\includegraphics[width=0.55\textwidth]{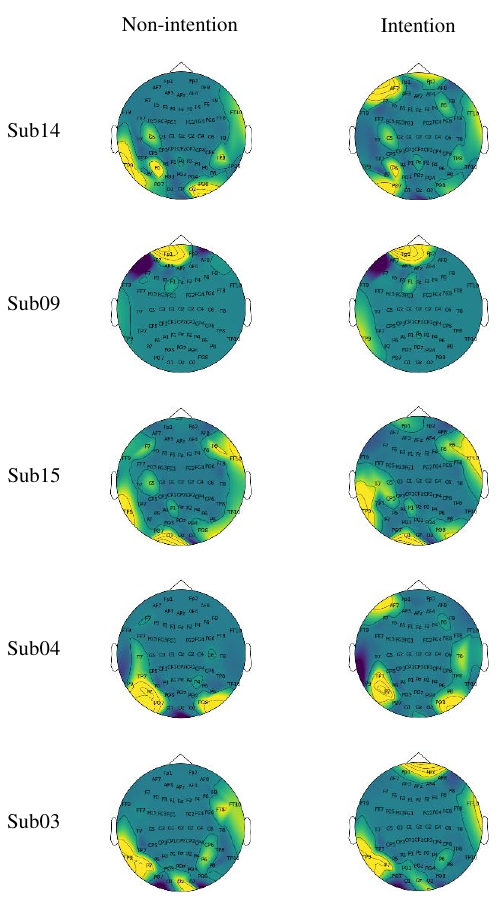}
\caption{The figure depicts the outcomes of the top-five participants with high classification results, as processed through the convolution block, both without (Non-intention) and with (Intention) the application of functional connectivity. Remarkably, in all five participants, a strong correlation was observed in the frontal and occipital areas under the Intention condition.}
\end{figure}

To quantitatively validate these findings, we compared the classification performance of the full FCDN model with that of the ablated model. Table 3.2 reports the classification accuracies for each subject in both settings. The model including the functional connectivity block achieved better performance in 14 out of 15 subjects, resulting in an average accuracy improvement of 2.37\%. These results demonstrate that although the base model without the functional connectivity block is already well aligned with the characteristics of complex visual imagery and captures relevant spatial-temporal features, the integration of visual cortex-guided functional connectivity further contributes to enhancing decoding performance by improving the spatial discriminability of neural representations.


In Fig. 3.3, we elucidate the efficacy of the functional connectivity block by contrasting the feature maps post convolution block, both with and without the inclusion of the functional connectivity block. These feature maps are visualized using topographical plots. For a robust evaluation, the results are presented specifically for the top 5 subjects who demonstrated superior classification performance when processed through the FCDN. Notably, in the feature maps corresponding to the `Intention' configuration (with the functional connectivity block), there is a discernible heightened activity in both the prefrontal and posterior regions, thereby underlining the pivotal role of the functional connectivity block. Conversely, the `Non-intention' configuration (absent the functional connectivity block) exhibits a diminished neural representation in these critical regions.


\section{Pseudo-Online Analysis}

\begin{figure*}[t!]
\centering
\includegraphics[width=\textwidth]{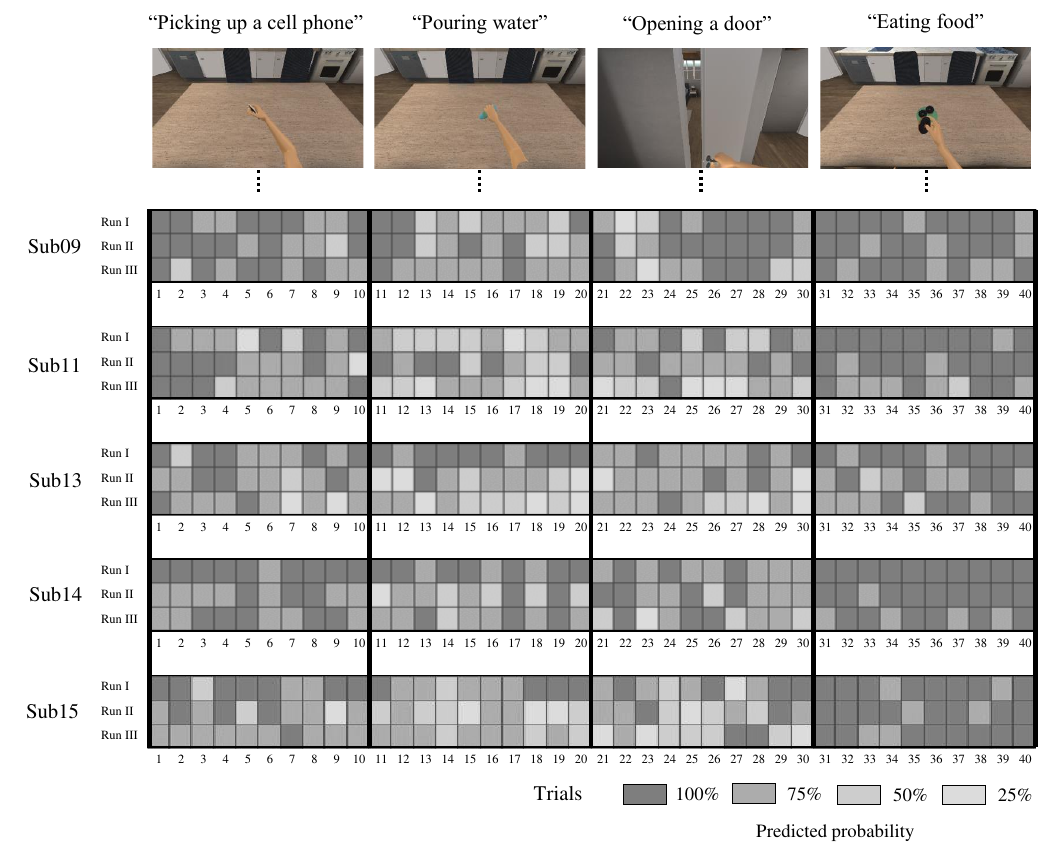}
\caption{The top three-dimensional virtual environment representations are the visual simulation on the monitor given to the user before the user performed complex visual imagery. Each of the four classes consists of 10 trials, expressed as follows: trial 1--10; picking up a cell phone (\textit{class-1}), trial 11--20; pouring water (\textit{class-2}), trial 21--30; opening a door (\textit{class-3}), and trial 31--40; eating food (\textit{class-4}). The figure below the visual stimulation is a representation of the classification outputs in the pseudo-online analysis. The darker grey squares indicate higher decoding success rates.}
\end{figure*}

We evaluated the proposed method through a pseudo-online analysis \cite{jeong2020brain, han2020enhanced} as depicted in Table 3.3. The EEG data were evaluated with a sliding window length of 2 s with a 50\% overlap. For each window, the results of the predicted class were derived, and the final prediction results of the trial were determined using the averages of the results of four windows. We derived the results by constructing the datasets into three online runs for accurate verification. Fig. 3.4 shows the pseudo-online performance for the top-five subjects. \textit{class-1} and \textit{class-4} generally recorded high success rates across the entire class, which are consistent with the results shown in Fig. 3.2. The results for the pseudo-online analysis of all subjects are shown in Supplementary Fig. 1. Table 3.3 displays the success rates of the pseudo-online analysis across the 15 subjects. An average performance exceeding 0.75 in four sliding windows has been the criterion for EEG-decoding success.

\begin{table}[t!]
\caption{Evaluation Performance for Pseudo-Online Analysis through the Success Rate of Decoding}
\renewcommand{\arraystretch}{0.9}
\small
\begin{center}
\resizebox{\textwidth}{!}{%
\begin{tabular}{ccccc}
\hline
\multirow{2}{*}{\textbf{Subjects}} & \multicolumn{4}{c}{\textbf{Success rate}}             \\ \cline{2-5} 
                                   & Run I       & Run II      & Run III     & Average     \\ \hline
Sub01                              & 0.73 (29/40) & 0.83 (33/40) & 0.58 (23/40) & 0.71 ($\pm$0.13) \\
Sub02                              & 0.63 (25/40) & 0.73 (29/40) & 0.63 (25/40) & 0.67 ($\pm$0.05) \\
Sub03                              & 0.73 (29/40) & 0.63 (25/40) & 0.73 (29/40) & 0.63 ($\pm$0.10) \\
Sub04                              & 0.80 (32/40) & 0.70 (28/40) & 0.80 (32/40) & 0.73 ($\pm$0.06) \\
Sub05                              & 0.63 (25/40) & 0.73 (29/40) & 0.63 (25/40) & 0.67 ($\pm$0.05) \\
Sub06                              & 0.70 (28/40) & 0.73 (29/40) & 0.70 (28/40) & 0.71 ($\pm$0.01) \\
Sub07                              & 0.73 (29/40) & 0.48 (19/40) & 0.73 (29/40) & 0.63 ($\pm$0.13) \\
Sub08                              & 0.68 (27/40) & 0.60 (24/40) & 0.68 (27/40) & 0.68 ($\pm$0.09) \\
Sub09                              & 0.90 (36/40) & 0.88 (35/40) & 0.90 (36/40) & 0.88 ($\pm$0.01) \\
Sub10                              & 0.70 (28/40) & 0.63 (25/40) & 0.70 (28/40) & 0.70 ($\pm$0.08) \\
Sub11                              & 0.58 (23/40) & 0.90 (36/40) & 0.63 (25/40) & 0.75 ($\pm$0.14) \\
Sub12                              & 0.60 (24/40) & 0.60 (24/40) & 0.65 (26/40) & 0.64 ($\pm$0.04) \\
Sub13                              & 0.63 (25/40) & 0.78 (31/40) & 0.65 (26/40) & 0.80 ($\pm$0.16) \\
Sub14                              & 0.80 (32/40) & 0.85 (34/40) & 0.88 (35/40) & 0.91 ($\pm$0.08) \\
Sub15                              & 0.73 (29/40) & 0.68 (27/40) & 0.75 (30/40) & 0.78 ($\pm$0.11) \\ \hline
\end{tabular}
}
\end{center}
\end{table}

Although we set this criterion strictly, with the proposed network, the average pseudo-online performance of all subjects across the three runs was 0.73. Among all subjects, Sub14 performed the best, scoring 0.91 over 40 trials.

\section{Classification Performance using Leave-One-Subject-Out Cross-Validation Approach}
We applied the LOSO approach \cite{lin2024eeg, fazli2009subject} to verify the possibility of solving subject-independent issues. In the LOSO method, the network is trained using the data from the source subjects and transferred to test the unknown data from the new subject. In this work, we set the training data to all subjects except one user who became a target. In LOSO conditions, baseline methods and proposed methods were trained using EEG data of 4 subjects, and a total of 16,000 training data were used, considering that the number of training data was 4,000 per subject. We selected the top five subjects who recorded high performance in the subject-dependent condition and confirmed the classification performance in the LOSO condition.

As shown in Table 3.4, the proposed network recorded the highest average classification performance in the LOSO condition compared to the baseline methods. In this study, the proposed method, FCDN, demonstrated a notable performance with a classification score of 0.4960, surpassing several established models. Specifically, it outperformed ConvNet (0.4498), EEGNet (0.4579), FBCNet (0.4598), and TSformer (0.4878). The classification performance of the proposed network in the LOSO condition was similar to that of the subject-dependent condition. These results suggest that the proposed network is robust under subject-independent conditions.

\begin{table}
\centering
\caption{Comparison of Classification Performance under LOSO Condition with Deep Learning Methods}
\renewcommand{\arraystretch}{1.1}
\small
\begin{tabular}{cccccc}
\hline
      & ConvNet                                                   & EEGNet                                                         & FBCNet                                                         & TSformer                                                  & FCDN                                                           \\ \hline
Sub04 & \begin{tabular}[c]{@{}c@{}}0.3813\\ ($\pm$0.0327)\end{tabular} & \begin{tabular}[c]{@{}c@{}}0.3968\\ ($\pm$0.0226)\end{tabular} & \begin{tabular}[c]{@{}c@{}}0.3845\\ ($\pm$0.0237)\end{tabular} & \begin{tabular}[c]{@{}c@{}}0.3987\\ ($\pm$0.0375)\end{tabular} & \begin{tabular}[c]{@{}c@{}}0.3950\\ ($\pm$0.0262)\end{tabular} \\
Sub06 & \begin{tabular}[c]{@{}c@{}}0.3313\\ ($\pm$0.0284)\end{tabular} & \begin{tabular}[c]{@{}c@{}}0.3252\\ ($\pm$0.0217)\end{tabular} & \begin{tabular}[c]{@{}c@{}}0.3472\\ ($\pm$0.0195)\end{tabular} & \begin{tabular}[c]{@{}c@{}}0.3886\\ ($\pm$0.0207)\end{tabular} & \begin{tabular}[c]{@{}c@{}}0.4013\\ ($\pm$0.0247)\end{tabular} \\
Sub09 & \begin{tabular}[c]{@{}c@{}}0.5138\\ ($\pm$0.0193)\end{tabular} & \begin{tabular}[c]{@{}c@{}}0.5125\\ ($\pm$0.0143)\end{tabular} & \begin{tabular}[c]{@{}c@{}}0.5320\\ ($\pm$0.0229)\end{tabular} & \begin{tabular}[c]{@{}c@{}}0.5103\\ ($\pm$0.0183)\end{tabular} & \begin{tabular}[c]{@{}c@{}}0.5225\\ ($\pm$0.0121)\end{tabular} \\
Sub14 & \begin{tabular}[c]{@{}c@{}}0.5825\\ ($\pm$0.0211)\end{tabular} & \begin{tabular}[c]{@{}c@{}}0.6375\\ ($\pm$0.0204)\end{tabular} & \begin{tabular}[c]{@{}c@{}}0.6067\\ ($\pm$0.0253)\end{tabular} & \begin{tabular}[c]{@{}c@{}}0.6592\\ ($\pm$0.0264)\end{tabular} & \begin{tabular}[c]{@{}c@{}}0.6588\\ ($\pm$0.0217)\end{tabular} \\
Sub15 & \begin{tabular}[c]{@{}c@{}}0.4400\\ ($\pm$0.0373)\end{tabular} & \begin{tabular}[c]{@{}c@{}}0.4175\\ ($\pm$0.0195)\end{tabular} & \begin{tabular}[c]{@{}c@{}}0.4287\\ ($\pm$0.0194)\end{tabular} & \begin{tabular}[c]{@{}c@{}}0.4823\\ ($\pm$0.0204)\end{tabular} & \begin{tabular}[c]{@{}c@{}}0.5025\\ ($\pm$0.0145)\end{tabular} \\
Avg.  & \begin{tabular}[c]{@{}c@{}}0.4498\\ ($\pm$0.0900)\end{tabular} & \begin{tabular}[c]{@{}c@{}}0.4579\\ ($\pm$0.2124)\end{tabular} & \begin{tabular}[c]{@{}c@{}}0.4598\\ ($\pm$0.0845)\end{tabular} & \begin{tabular}[c]{@{}c@{}}0.4878\\ ($\pm$0.0782)\end{tabular} & \begin{tabular}[c]{@{}c@{}}0.4960\\ ($\pm$0.0964)\end{tabular} \\ \hline
\end{tabular}
\end{table}

\chapter{Discussion}\label{chap:discussion}
Complex visual imagery provides a paradigm by which users can effectively imagine an intended movement by only thinking. We identified the neurophysiological features of complex visual imagery and proposed a network to investigate the feasibility of BCI applications. We have highlighted the possibility of identifying users' intentions in real-time via pseudo-online analysis and demonstrated the possibility of developing a subject-independent system via LOSO.

We have provided a highly intuitive BCI paradigm for subjects to improve EEG signal decoding performance for the development of BCI applications. As a BCI paradigm, we provided complex visual imagery to subjects in this study, and then performed data analysis to verify the quality of the EEG data obtained using our paradigm. The measurement of power spectra in the channels responsible for each brain region showed significant results in the prefrontal and occipital regions, indicating a similar tendency to conventional visual imagery-related studies \cite{sousa2017pure, koizumi2019eeg, xie2020visual, llorella2021classification, kilmarx2024evaluating}.

We also compared the classification performance of the proposed network with that of conventional decoding approaches. The proposed network recorded a higher performance for the 4-class condition than did the conventional methods, and we deduced that the proposed network is an effective network for decoding complex visual imagery. We observed a consistent performance trend across subjects for the various deep learning networks, including the proposed model. However, this trend was not aligned with the performance pattern observed in the FBCSP approach, indicating that the subject-wise classification tendencies differ significantly between FBCSP and deep learning-based methods. For visual imagery, both spatial and temporal information is important. Consequently, the FBCSP approach, which does not emphasize temporal information, demonstrated lower performance and the constant tendency among subjects was eliminated.

As complex visual imagery data are based on the user's imagined temporal changes in a 3D space, we designed the proposed network accordingly. The network proposed in this study emphasized channels containing meaningful features related to complex visual imagery based on PLVs and used these to modify the raw EEG data. In addition, the channel-wise separable convolutional layer was placed on the last layer of the CNN. Features were extracted using the two preceding convolutional layers to include temporal information, and these features were used as inputs for DeiT. This structure extracted information based on features, including temporal information, when extracting the spatial features of the input using DeiT. Thus, the output contains both spatial and temporal information.

\begin{table}[]
\centering
\caption{The Pearson Correlation Coefficient Results for Phase-Locking Value in Representative Subjects}
\renewcommand{\arraystretch}{1.0}
\begin{tabular}{cccccc}
\hline
\multicolumn{6}{c}{\textit{r}-vlaue}                        \\ \hline
      & Sub04 & Sub06  & Sub09  & Sub14  & Sub15   \\ \hline
Sub04 & 1.0000     & 0.7881 & 0.8555 & 0.7390 & 0.4025 \\
Sub06 & -     & 1.0000      & 0.5627 & 0.6523 & 0.4864  \\
Sub09 & -     & -      & 1.0000      & 0.6612 & 0.2755 \\
Sub14 & -     & -      & -      & 1.0000      & 0.4495  \\
Sub15 & -     & -      & -      & -      & 1.0000       \\ \hline
\end{tabular}
\end{table}

In the pseudo-online analysis, the final decisions were derived using four segmented windows, which yielded a high decoding performance. This confirms that complex visual imagery can be beneficial in real-life applications for users, and demonstrates the effectiveness of the proposed network in decoding complex visual imagery. Notably, this technology shows potential in aiding individuals through BCI-based applications, such as robotic arms, suggesting a promising avenue for practical implementation. While this study was conducted with a relatively small sample size of 15 participants, consistent decoding patterns were observed across subjects. The robust classification performance under the LOSO validation scheme suggests that the proposed method captures generalizable neural features across individuals.

For all subjects, all deep learning approaches yielded a classification performance of $\geq$ 0.25, which is the chance level. The proposed network yielded the highest level of performance. The complex visual imagery is possible in a subject-independent approach because it demonstrated a constant data distribution, regardless of the subject, and the spatial area emphasized in the network was similar. 

As shown in Table 4.1, the similarity of PLVs among the subjects was calculated using Pearson's correlation coefficient (CC) \cite{benesty2009pearson} $r$, which refers to the similarity of PLVs among subjects, and were mostly above 0.3, with a distinct amount of correlation. The CCs of all subjects are shown in Supplementary Table III. As most users have PLVs with distinct correlations, we could infer that users were imagining in a similar manner, which could be the starting point for the development of a subject-independent approach. All three cases ($r$ $\leq$ 0.3) were related to Sub15, indicating that Sub15 used a different pattern of imagination from other subjects. If re-learning is requested from these types of subjects or a deep learning approach that minimizes spatial information is applied, better performance could be expected. In this study, we constructed a network in which spatial information was emphasized based on PLVs, which can be used as an indicator of data learning in subject-independent conditions.

\chapter{Conclusion}\label{chap:conclusion}
In this study, we applied a 3D-BCI training platform allowing users to perform complex visual imagery effectively. Using the 3D-BCI training platform, we were able to collect high-quality visual imagery data from participants, and we verified the quality of the data using neurophysiological methods. In addition, we proposed FCDN to decode the obtained visual imagery data. The network was designed to consider both spatial and temporal information based on the characteristics of visual imagery. Using functional connectivity, we emphasized channels that were highly correlated with visual imagery, and accordingly extracted spatial information using convolutional layers. Subsequently, we used the extracted features as inputs to DeiT. The network proposed in this study recorded a robust decoding performance regardless of the number of classes. Achieving high classification performance for various classes is important, but it is also significant that the types of classes proposed in this study were designed to involve complex upper-limb movements.
While the current study involved 15 participants, the consistent subject-wise decoding performance suggests that the proposed method captures generalizable neural representations. Nevertheless, we recognize the importance of further validation with a broader population. To this end, we plan to conduct additional experiments to collect data from more participants, aiming to enhance the generalizability and robustness of the proposed method. We will also further develop FCDN, the proposed network, for high classification performance, so that it can be applied to real-time BCI scenarios. In addition, we will explore methods to reduce the computational complexity of the networks proposed in this study. We also aim to expand the dataset by increasing the number of movement classes, allowing the model to decode a broader range of upper-limb movements. Based on this, we will apply the proposed network to an EEG-based robotic arm system, which will contribute to the development of rehabilitation systems for patients with tetraplegia or for supporting the daily life activities of healthy people.\\

\newpage
\renewcommand\bibname{Reference}
\bibliographystyle{ieeetr}
\bibliography{REFERENCE}

\end{document}